\documentclass[aps,pra,showpacs,twoside,10pt,floatfix,nofootinbib,longbibliography,twocolumn]{revtex4-2}
\usepackage[colorlinks=true, citecolor=red, urlcolor=blue]{hyperref}
\usepackage{amssymb,amsfonts,amsmath,bm,amsthm,mathtools,braket,geometry,xcolor,times,comment}
\usepackage[normalem]{ulem}
\begin{document}

\title{Scaling of entanglement entropy and correlations in the variable-range  extended Ising model}
\author{Harikrishnan K J$^1$, Debasis Sadhukhan$ ^{2,3}$, Amit Kumar Pal$^1$}
\affiliation{$^1$ Department of Physics, Indian Institute of Technology Palakkad, Palakkad 678 623, India \\
$^2$ Department of Physics, Institute of Science, Banaras Hindu University, Varanasi 221 005, India \\
$^3$ Department of Physics, Indian Institute of Engineering Science and Technology, Shibpur, Howrah 711103, India}

\begin{abstract}

We study the two-point correlation functions and the bipartite entanglement in the ground state of the exactly-solvable variable-range extended Ising model of qubits in the presence of a transverse field on a one-dimensional lattice. We introduce the variation in the range of interaction by varying the coordination number, $\mathcal{Z}$, of each qubit, where the interaction strength between a pair of qubits at a distance $r$ varies as $\sim r^{-\alpha}$. We show that the algebraic nature of the correlation functions is present only up to $r=\mathcal{Z}$, above which it exhibits short-range exponential scaling. We also show that at the critical point, the bipartite entanglement exhibits a power-law decrease ($\sim\mathcal{Z}^{-\gamma}$) with increasing coordination number irrespective of the partition size and the value of $\alpha$ for $\alpha>1$. We further consider a sudden quench of the system starting from the ground state of the infinite-field limit of the system Hamiltonian via turning on the critical Hamiltonian, and demonstrate that the long-time averaged bipartite entanglement exhibits a qualitatively similar variation ($\sim\mathcal{Z}^{-\gamma}$) with $\mathcal{Z}$.       

\end{abstract}

\maketitle

\section{Introduction} 
\label{sec:intro}

Long-range systems are not only abundant in nature \cite{RevModPhys.95.035002, Padmanabhan, Chavanis,LandauLifshitz,dipolExp1, dipolExp2,spinice1, spinice2}, but also inherent in numerous experimental testbeds~\cite{RevModPhys.82.2313, RevModPhys.93.025001, Hall2016-ld, Davis2023-ca, PhysRevLett.88.067901}.
They have gained much attention over the last decade due to their counter-intuitive  properties~\cite{Richerme2014,PhysRevB.93.125128,1702.05368,Ares19,PhysRevA.97.062301, PhysRevLett.17.1133, PhysRev.158.383, PhysRevLett.109.025303, PhysRevX.3.031015, PhysRevB.87.035114, RevModPhys.82.277, PhysRevLett.109.267203,PhysRevLett.119.170503} as opposed to the conventional expectations derived from traditional statistical mechanics~\cite{Cardy1996-tq,Mussardo2020-hp, Sachdev2011-jw}. This is primarily because statistical mechanics often formulate a universal theory starting from a short-range toy model~\cite{Ma1985-dy,Ma2019-gt}, or a local $\phi^4$-field theory~\cite{Sachdev2011-jw}. Thanks to the universality principle~\cite{RevModPhys.70.653}, various physical scenarios near criticality can be described using a unified theoretical framework of continuous quantum phase transitions (QPTs)~\cite{Sachdev2011-jw}.  Here, the onset of a QPT is usually associated with a diverging correlation length~\cite{Sachdev2011-jw, Nishimori2010-yd}, which stems from an algebraically decaying correlation function. On the other hand, away from the critical point, correlation function, $C_r$, decays exponentially with distance, $r$, as  $C_r \sim e^{-r/\xi}$, exhibiting a finite length scale of the correlation length $\xi$~\cite{Fisher1964-gw}. This conventional narrative of correlation length falls short when one tries to extrapolate it to the long-range regime~\cite{RevModPhys.95.035002}. 

Correlation functions of longer-range systems can manifest persistent algebraic tails regardless of the proximity to the critical point~\cite{PhysRevLett.29.917,Vodola2014,Vodola2015,Debasis2021,PhysRevA.98.023607,PhysRevB.96.104432,PhysRevLett.111.207202,PhysRevB.94.075156,PhysRevLett.111.260401}. In these systems, therefore, correlation length derived from exponential fit loses its steadfast connection to critical phenomena, making it difficult to place within the conventional framework of continuous QPT~\cite{Sachdev2011-jw,Fisher1964-gw}. This highlights the need of a new framework to accommodate the notion of correlation length in long-range systems. There has been a significant body of work~\cite{PhysRevB.8.281,Honkonen1990-kz,PhysRevE.92.052113,PhysRevLett.87.137203,PhysRevLett.118.241601,PhysRevE.95.012143,PhysRevE.96.012108,Lepori2017-fd,Brankov1992-ol,Debasis2021,SciPostPhys.13.4.088,PhysRevLett.130.246601,PhysRevB.108.165130,Vodola2015,PhysRevA.98.023607} to understand how this transition occurs, particularly focusing on how the increase in the range of interactions alters the physical properties. Typically, this is done by reducing the exponent $\alpha$ of a power-law interaction, $J(r) \sim 1/r^{\alpha}$, between subsystems at a distance $r$. For $\alpha$ greater than a model-dependent critical value $\alpha_c$, the system belongs to the short-range universality class~\cite{PhysRevB.8.281,PhysRevB.96.104432}. For $\alpha<D$, referred to as the strong long-range regime~\cite{RevModPhys.95.035002} with $D$ being the lattice dimension,  the energy becomes extensive leading to difficulty in thermodynamic-limit calculation~\cite{Campa2014-rl}. However, in the weak long-range regime~\cite{RevModPhys.95.035002} $D< \alpha < \alpha_c$,  thermodynamics remains well-defined, yet universal properties are influenced by long-range interactions. Therefore, varying $\alpha\in[0,\infty]$  offers the possibility to explore the physics for the short-range to long-range transition.

When the range of interaction in an exactly-solvable one-dimensional ($D=1$) qubit system~\cite{Vodola2014,Vodola2015,Debasis2021,Ganesh2022} is probed via the exponent $\alpha$ using the scaling of the \emph{classical} correlation functions, three different scalings  emerge in the infinite-range ($\mathcal{Z}\rightarrow\infty$) model, separated by the points $\alpha=D=1$ and $\alpha=\alpha_c=2$~\cite{RevModPhys.95.035002}. The nature of the correlation for $\alpha>2$ effectively resembles the short-range model~\cite{RevModPhys.95.035002} with correlations exhibiting exponential tail except at the critical point, where the correlations are algebraic \cite{Vodola2015, Debasis2021}. The long-range interactions start influencing the correlation functions for $\alpha<2$ by introducing an algebraic tail. In the weak long-range regime $1<\alpha<2$, the correlations at short distances are dominated by an exponential contribution, which, in the strong long-range regime ($\alpha<1$), becomes negligible \cite{Vodola2015}.

In similar models, quantum correlations belonging to the entanglement-separability paradigm~\cite{RevModPhys.81.865} -- specifically bipartite entanglement between a block of $M$ qubits with the rest of the system, as quantified by the von Neumann entropy~\cite{PhysRevA.54.3824,RevModPhys.81.865} denoted by $S_M$ --  also show different scaling behavior depending on different values of $\alpha$~\cite{PhysRevA.97.062301, Solfanelli2023-nq,Ganesh2022}.  In the $\alpha>2$ regime, entanglement follows the area law~\cite{PhysRevLett.71.666, RevModPhys.82.277}, i.e., grows as the boundary of the considered subsystem, i.e., $S_{M}\sim M^{D-1}$, when the system is gapped i.e.\ far from the critical point. For one-dimensional short-range gapped systems, it was shown that $S_M$ is constant~\cite{Hastings2007-zr}, following the area law. At the critical point, however, the area law is known to be violated by a logarithmic term as $S_M\sim c_0 \log M$, where $c_0$ is the central charge coming from conformal field theory~\cite{Calabrese2004}. In the weak long-range regime ($1<\alpha<2$), the system follows the area law  and  exhibits standard logarithmic deviations from the entanglement area law at the critical point, although the coefficients in front of these logarithmic divergences in this regime differs from the prediction of critical conformal field theory~\cite{Calabrese2004, Calabrese2009-lx}. There is also another sub-leading contribution, which makes the accurate scaling to be $S_M = c_0' \log M + c_1 M^{-\delta}+c_2$, where $c_0', c_1, c_2$ and $\delta$ are constants depending on the system parameters \cite{Solfanelli2023-nq}. The picture becomes more involved in the strong long-range regime ($\alpha<1$), where the system shows genuine non-additivity, exhibiting a logarithmic deviation from the area law even away from criticality~\cite{1506.06665, PhysRevA.97.062301, Solfanelli2023-nq}. Certain specific situations can also allow long-range interactions to exhibit sub-logarithmic growth of entanglement entropy \cite{1405.2804,1611.08506,1804.06357,Ares19, Solfanelli2023-nq}, and can even manifest as a volume law~\cite{1401.5922}.

The transition from short-range to long-range system can also be carried out by increasing the range of interaction. Specifically, for a fixed $\alpha$, one can increase the range by increasing the coordination number (c.f.~\cite{Ganesh2022}), $\cal{Z}$, and refer to such situation as the \emph{variable-range} (VR) interactions. Thanks to recent progress in experimental quantum technologies, VR interactions has been realized in Rydberg gases~\cite{Bottcher2021-ge}, and simulated using a programmable quantum computer~\cite{Matos2023-us, PhysRevResearch.6.013311}. Such VR interactions have also been shown to affect topological properties~\cite{PhysRevB.95.195160,Kartik2021,Kartik2024}. However, a systematic analysis of the classical correlation functions and bipartite entanglement with varying $\mathcal{Z}$ remains mostly unexplored.

In this work, we consider an exactly-solvable qubit-system with $D=1$ and VR interactions, and investigate the trends of classical correlations and bipartite entanglement present in the system while crossing over to the  long-range  regime from the short-range regime by varying $\cal{Z}$. We diagonalize the Hamiltonian, and perform analytical calculations at the thermodynamic limit, setting $\alpha>1$. A leading order approximation in $\mathcal{Z}$ reveals the emergence of $r=\mathcal{Z}$ as a natural length scale for the transition from effective short-range to long-range behavior of the system. We demonstrate that for distances $r<\mathcal{Z}$, correlation functions always decay algebraically, indicating an effective long-range behavior. On the other hand, when $r>\mathcal{Z}$, effective short-range behavior is present in the system, manifested by the occurrence of algebraic as well as exponential decay of the correlation functions. Further, we derive a $\mathcal{Z}$-dependent quasi-particle velocity that flags the weak long-range and effective short-range regimes on the $\alpha$ axis.

We also investigate how entanglement scales with  $\cal{Z}$, keeping $\alpha$ and $M$ to be fixed. For the  ``$1$:rest" bipartition  ($M=1$),  we analytically show, at the critical point, that for $\alpha=2$, $S_M\sim\mathcal{Z}^{-1}$, while for $\alpha>1$, $S_M\sim \mathcal{Z}^{-\gamma}$.  We numerically show that the $\alpha$- dependence of $\gamma$ is quadratic in nature, and the results remain qualitatively unchanged for all values of $M>1$. Further, starting from the ground state of the Hamiltonian at the infinite-field limit, and choosing a simple sudden quench via the critical Hamiltonian, we show that the large-time average of bipartite entanglement also exhibits similar power-law variation with $\mathcal{Z}$. 

The rest of the paper is organised as follows: In Sec.~\ref{sec:model_and_diagonalization}, we introduce our prototypical model and give a brief summery of its diagonalization procedure. The static results concerning bipartite entanglement are given in Sec.~\ref{sec:biartite_entanglement} while the results of the quench dynamics are contained in Sec.~\ref{sec:dynamics}. Finally we conclude in Sec.~\ref{sec:conclusion}. 

\begin{figure}
    \centering
    \includegraphics[width=0.8\linewidth]{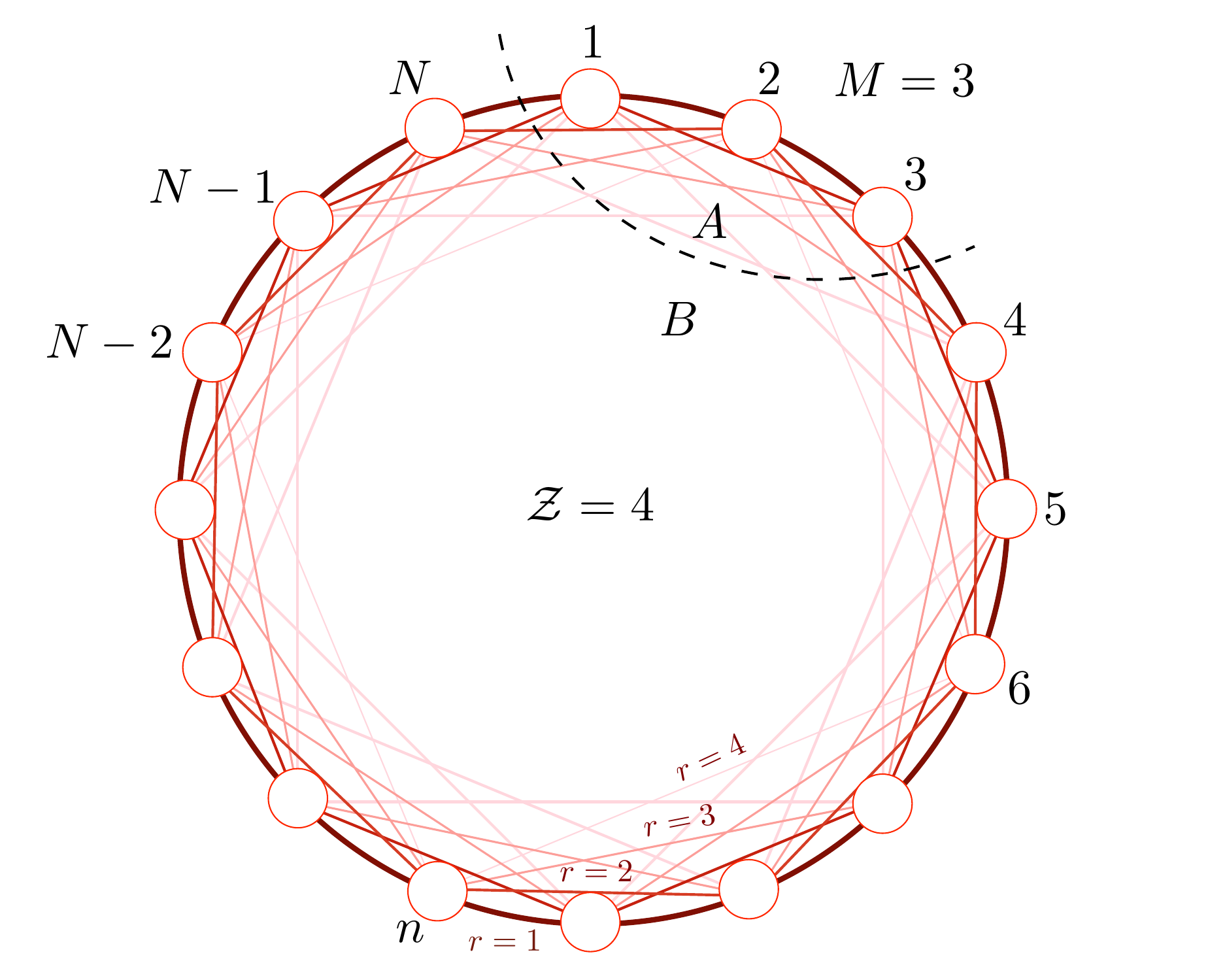}
    \caption{A chain of $N$ interacting qubits with periodic boundary condition described by the Hamiltonian~(\ref{eq:full_hamiltonian}), where each qubit $n$ has a coordination number $\mathcal{Z}=4$. The strength of the interactions between the qubits at a distance $r=1,2,3,4$ (see the $n$th qubit, where different lines represent interactions with qubits at different distances) are represented by the different shades of the lines -- darker shade implying a higher interaction strength due to a lower distance for a fixed $\alpha$ (see Eq.~(\ref{eq:Jr})). We compute the bipartite entanglement between an $M$-qubit block (for example, $M=3$ in the figure), representing the party $A$, and the rest of the qubits, representing the party $B$, as quantified by the von Neumann entropy as a function of $\mathcal{Z}$.}
    \label{fig:schematic}
\end{figure}

\section{Variable-Range Extended Ising Model}
\label{sec:model_and_diagonalization}

We consider the variable-range extended Ising (VREI) model on a system of $N$ qubits arranged on a one-dimensional (1D) lattice with periodic boundary condition (PBC), described by the Hamiltonian~\cite{Ganesh2022} (see Fig.~\ref{fig:schematic})
\begin{equation}\label{eq:full_hamiltonian}
    H=-\sum_{n=1}^N\sum_{r=1}^\mathcal{Z} J_r \sigma_n^x\sigma_{n+r}^x\prod_{m=n+1}^{n+r-1} \sigma_m^z - \frac{h}{2} \sum_{n=1}^N\sigma_n^z,
\end{equation}
where $n$ is the qubit-index, $J_r$ is the strength of the ferromagnetic (FM)~\cite{Ganesh2022,Vodola2014,Vodola2015}  interaction ($J_r>0$) between two qubits at a distance $r$, $h$ is the  strength of the \emph{transverse} qubit-local magnetic field, and $\mathcal{Z}$ is the \emph{coordination number} representing the number of qubits connected to each qubit via the FM interaction. 
Further, we assume the FM interaction strength $J_r$ decreasing with $r$ following a \emph{power law}, given by~\cite{Ganesh2022,Vodola2014,Vodola2015} 
\begin{eqnarray}\label{eq:Jr}
    J_r=r^{-\alpha}/A,
\end{eqnarray}
where $A=\sum_{r=1}^\mathcal{Z}r^{-\alpha}$ is the Kac normalization constant~\cite{Kac1963-pi} ensuring $\sum_r  J_r=1$. 

The parity symmetry, $[H,P]=0$ with  $P=\prod_{n=1}^N \sigma_n^z$, splits the Hamiltonian into positive and negative parity sectors as 
\begin{equation}
    H=P^+H^+P^++P^-H^-P^-,
\end{equation}
with $P^{\pm}=[1\pm P]/2$. The Jordan-Wigner transformation~\cite{Glen2024}, given by 
\begin{eqnarray}
    \sigma^x_n &=& -(c_n+c_n^\dagger)\prod_{m<n} \sigma_m^z, \\
    \sigma^y_n &=& i(c_n-c_n^\dagger)\prod_{m<n} \sigma_m^z, \\
    \sigma^z_n &=& 1-2c_n^\dagger c_n,
    \label{eq:jw_transformation}
\end{eqnarray}
leads to the \emph{full} fermionic Hamiltonian
\begin{eqnarray}\label{eq:fermionized}
    H&=&-\sum_{n=1}^N\sum_{r=1}^{\mathcal{Z}} J_r \left( c_n^\dagger c_{n+r}^\dagger + c_n^\dagger c_{n+r} + h.c. \right)\nonumber\\&& -\frac{h}{2}\sum_{n=1}^N \left(1-2c_n^\dagger c_n \right). 
\end{eqnarray}
while $H^+(H^-)$, in fermionic description, correspond to the sectors having even (odd) number of fermions with anti-periodic (periodic) boundary condition, given by $c_{n+N}=-c_n$ ($c_{n+N}=c_n$). In the next step, the Hamiltonian is taken through a Fourier transformation of the fermionic operators, given by 
\begin{eqnarray}
    c_n &=& \frac{e^{-i\pi/4}}{\sqrt{N}}\sum_k c_k e^{ikn}, 
\end{eqnarray}
and the calculation is restricted to $H^+$ with $N$ chosen to be even, as the ground state is always obtained from the even fermion sector~\cite{Glen2024}. Application of the Fourier transformation and subsequently implementing the anti-periodic boundary condition by choosing the quasi-momenta $k$ from the set
\begin{eqnarray}
    \mathcal{K}^+&=&\bigg\{\pm(2q-1)\frac{\pi}{N};q=1,2,3\cdots\frac{N}{2}\bigg\},\label{eq:ABC_k}
\end{eqnarray}
$H^+$ takes the form 
\begin{eqnarray}
    H^+= 2\sum_{k>0} \begin{bmatrix}
        c_k &
        c_{-k}^\dagger
    \end{bmatrix}
    H_k^+
    \begin{bmatrix}
        c_k \\
        c_{-k}^\dagger
    \end{bmatrix},\label{eq:H+_intermediate}
\end{eqnarray}
with 
\begin{eqnarray}
    H_k^+&=&  \left[h/2 - \text{Re}(\tilde{J}_k)\right]\sigma^z +  \text{Im}(\tilde{J}_k)\sigma^x 
\end{eqnarray}
being a $2\times 2$ matrix,  and 
\begin{eqnarray}\label{eq:Jk_tilde}
    \tilde{J}_k = A^{-1}\sum_{r=1}^{\mathcal{Z}} r^{-\alpha} e^{ikr}.
\end{eqnarray}
We now perform a Bogoliubov transformation~\cite{Glen2024} defining the new quasi particles as 
\begin{eqnarray}
    \gamma_k &=& U_k c_k + V_k c_{-k}^\dagger\\
    \gamma_{-k} &=& -V_k^* c_k^\dagger + U_k^* c_{-k}
\end{eqnarray}
with  
\begin{eqnarray}\label{eq:eigenstates}
    U_k &\sim& \frac{h}{2}-\text{Re}(\tilde{J}_k)+\left[\frac{h^2}{4}-h\text{Re}(\tilde{J}_k)+|\tilde{J}_k|^2\right]^{\frac{1}{2}},\nonumber\\
    V_k&\sim& \text{Im}(\tilde{J}_k),
\end{eqnarray}
up to normalization. This  diagonalizes $H_k^+$, and subsequently $H^+$ as
\begin{eqnarray}
    H^+ &=& \sum_{k} \left(\gamma_k^\dagger \gamma_k-\frac{1}{2}\right) \omega_k\label{eq:final_Hk}    
\end{eqnarray}
providing the energy eigenvalues $\pm\omega_k$ with 
\begin{eqnarray}\label{eq:dispersion}
    \omega_k=2\sqrt{\left[h/2 - \text{Re}(\tilde{J}_k)\right]^2 +  \left[\text{Im}(\tilde{J}_k)\right]^2}. 
\end{eqnarray}
The ground state of the Hamiltonian is the Bogoliubov vacuum state with $\gamma_k^\dagger\gamma_k=0\forall k$, with the ground state energy $E_0=-\frac{1}{2}\sum_{k}\omega_k=-\sum_{k>0}\omega_k$.

\subsection{Effective short-range behavior in small \texorpdfstring{$k\mathcal{Z}$}{kZ} limit}
\label{subsec:short_range}

\begin{figure*}
    \centering
    \includegraphics[width=0.7\linewidth]{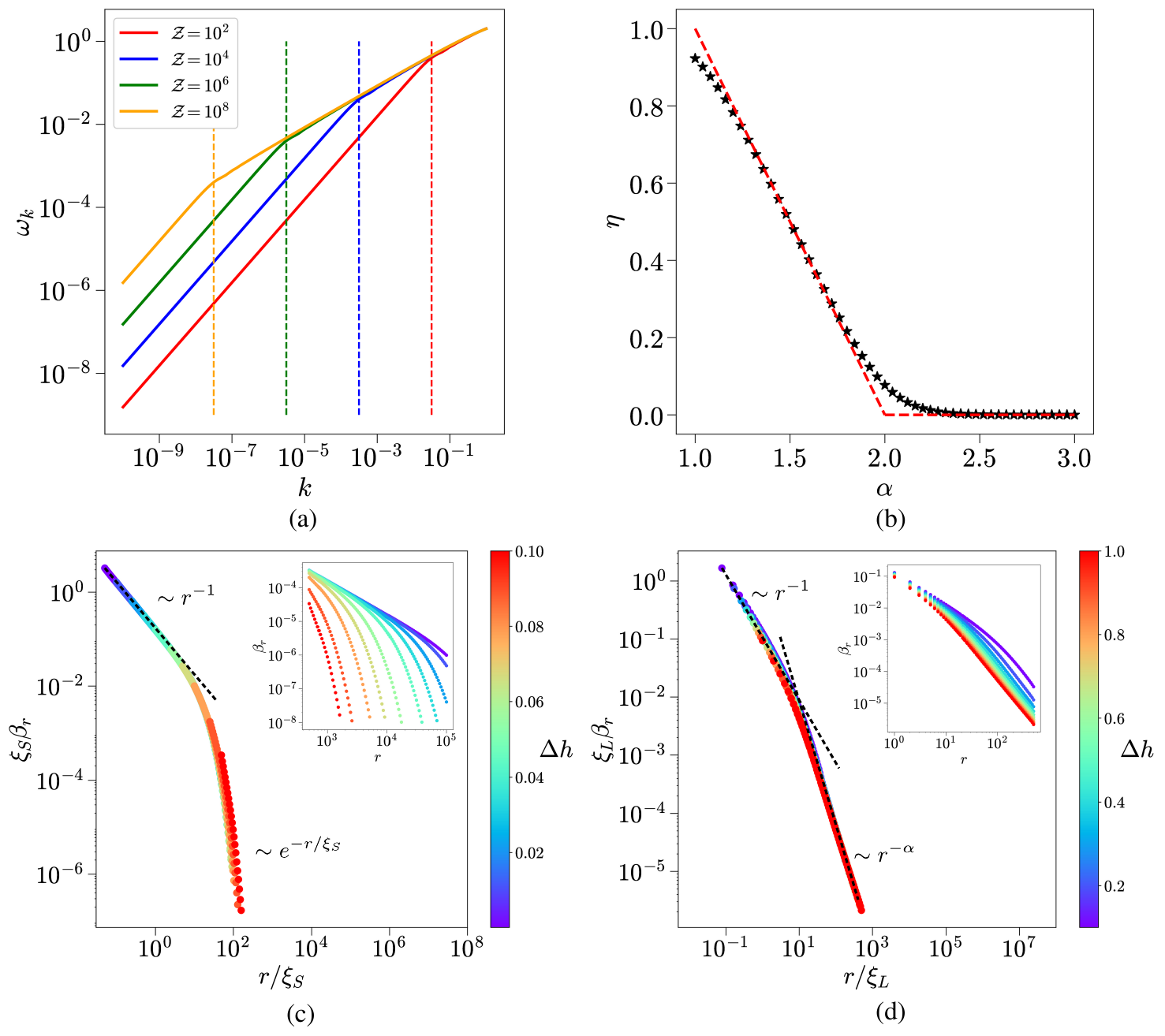}
    \caption{ (a) Dispersion relation (see Eq.~(\ref{eq:dispersion})) in the log-log scale at the critical point $h_c=2$ for different  $\mathcal{Z}$, computed by exact numerical evaluation of $\tilde J_k$ for $\alpha=1.5$. The crossover points $k=k_\mathcal{Z}$  in the Brillouin zone (see Eq.~(\ref{eq:kz})) are shown by vertical dashed line for different $\mathcal{Z}$. (b) Variation of the exponent $\eta$ (see Eq.~(\ref{eq:qp_velocity_largeZ})) against $\alpha$ at the critical point, obtained by a numerical fit of $\log(d\omega_k/dk)$ against $\log\mathcal{Z}$ in the range $4\leq \log\mathcal{Z}\leq 7$, which matches Eq.~(\ref{eq:eta}) shown by the red dashed line. (c) Data collapse via appropriate scaling of the two point correlation function $\beta_r$ in the short-range regime $r>\mathcal{Z}$ with $\mathcal{Z}=500$ and $\alpha=1.9$. The $r\ll\xi_L$ regime shows a decay $\beta_r\sim r^{-1}$, while $r\gg\xi_L$ exhibits $\beta_r\sim e^{-r/\xi_S}$. (d) Scaling of $\beta_r$ in the long-range regime $r<\mathcal{Z}$, where the parameters are kept as in (c). The insets of both (c) and (d) depict the variation of $\beta_r$ with $r$. The $r\ll\xi_L$ regime shows an algebraic decay $\beta_r\sim r^{-1}$, while an exponential decay $\beta_r\sim r^{-\alpha}$ is observed in $r\gg\xi_L$. All quantities plotted are dimensionless.}
    \label{fig:dispersion}
\end{figure*}

The normalization constant $A$ (see Eq.~(\ref{eq:Jr})) can be approximated as
\begin{eqnarray}\label{eq:A}
    A&=&\sum_{r=1}^\mathcal{Z}r^{-\alpha} 
    \approx \zeta(\alpha) + \frac{\mathcal{Z}^{1-\alpha}}{1-\alpha}, 
\end{eqnarray}
where $\zeta(\alpha)=\sum_{r=1}^\infty r^{-\alpha}$ is the Riemann zeta function, which is convergent for $\alpha>1$ (see Appendix~\ref{app:calculation} for details). Further,
\begin{eqnarray}
    \sum_{r=1}^\mathcal{Z}e^{ikr}r^{-\alpha}     &\approx& \mathbf{Li}_\alpha(e^{ik})-\int_{r=\mathcal{Z}}^\infty e^{ikr}r^{-\alpha}dr, 
\end{eqnarray}
where $\mathbf{Li}_\alpha$ is the polylogarithmic function. Using the asymptotic form of polylogarithmic function~\cite{Debasis2021}, and converting the integral into a lower incomplete Gamma function $\gamma(s,x)=\int_0^x e^{-t} t^{s-1}dt$ (see Appendix~\ref{app:calculation}), one arrives at 
\begin{eqnarray}
    \sum_{r=1}^\mathcal{Z}e^{ikr}r^{-\alpha}
    &\approx&(-ik)^{\alpha-1}\gamma(1-\alpha,-ik\mathcal{Z})\nonumber\\&&+ \sum_{n=0}^\infty \frac{\zeta(\alpha-n)}{n!}(ik)^n.
\end{eqnarray}
In the limit $k\mathcal{Z}\ll 1$, $\gamma(1-\alpha,-ik\mathcal{Z})$ can be approximated, up to the first order in $k$, as
\begin{eqnarray}\label{eq:incomplete_gamma_simplified_first_order}
    \gamma(1-\alpha,-ik\mathcal{Z}) &=&
    (-ik)^{1-\alpha}\bigg[\frac{\mathcal{Z}^{1-\alpha}}{1-\alpha}+ik\frac{\mathcal{Z}^{2-\alpha}}{2-\alpha}\bigg].
\end{eqnarray}
Using Eqs.~(\ref{eq:A}) and (\ref{eq:incomplete_gamma_simplified_first_order}) in Eq.~(\ref{eq:Jk_tilde}), and subsequently retaining the terms up to the first order in $k$, one obtains  
\begin{eqnarray}\label{eq:shortrangeReIm}
     \text{Re}(\tilde{J}_k)
    &=& 1, \text{ and }
     \text{Im}(\tilde{J}_k)
    = kv_{\alpha,\mathcal{Z}},
\end{eqnarray}
where 
\begin{equation}\label{eq:qp_velocity}
v_{\alpha,\mathcal{Z}}=\frac{\zeta(\alpha-1)+(2-\alpha)^{-1}\mathcal{Z}^{2-\alpha}}{\zeta(\alpha)+(1-\alpha)^{-1}\mathcal{Z}^{1-\alpha}}.  
\end{equation}
is the $\alpha$ and $\mathcal{Z}$ dependent quasi-particle velocity $v_{\alpha,\mathcal{Z}}=d\omega_k/dk$ at the critical point $h_c=2$\footnote{Note that in the $\mathcal{Z}\rightarrow\infty$ limit of the VREI model, two critical points exist, given by (a) $h^{(1)}_c=2$, for which the gap closes at $k=0$,  and (b) $h_c^{(2)}=-2(1-2^{1-\alpha})$, for which the gap closing happens at $k=\pi$. For an arbitrary but finite $\mathcal{Z}$, $h_{c}^{(1)}$ remains unchanged, while $h_c^{(2)}$  shifts with varying $\mathcal{Z}$~\cite{Ganesh2022}. Throughout the paper, unless otherwise stated, we look at the fixed critical point $h_c=h_c^{(1)}$, and its neighborhood.}. Using these results in Eq.~(\ref{eq:dispersion}) leads to the expression for $\omega_k$ as
\begin{eqnarray}
    \omega_k &=& 2\sqrt{\left(h/2-1\right)^2 + k^2v_{\alpha,\mathcal{Z}}^2},
\end{eqnarray}
which, at the critical point $h_c$, becomes $\omega_k=2kv_{\alpha,\mathcal{Z}}$. For arbitrary yet fixed values of $\alpha(>1)$ and $\mathcal{Z}$, $\omega_k\propto k$, implying a short-range behavior of the VREI model. 

For $\mathcal{Z}=1$, $H$ (Eq.~(\ref{eq:full_hamiltonian})) represents a \emph{truly} short-range model in the entire Brillouin zone $[-\pi,\pi]$, irrespective of the value of $\alpha$. On the other hand, for $\mathcal{Z}>1$, the length scale $\mathcal{Z}^{-1}$ decides the behavior of the VREI model in the Brillouin zone independent of all $\alpha>1$.
The short-range behavior of the model is observed for  
\begin{eqnarray}\label{eq:kz}
    k<k_{\mathcal{Z}}=\pi\mathcal{Z}^{-1}. 
\end{eqnarray}
This is in contrast to the limit $k\mathcal{Z}\gg 1$ considered in~\cite{Debasis2021}, resulting in $\gamma(1-\alpha,-ik\mathcal{Z})\rightarrow \Gamma(1-\alpha)$ -- the Gamma function -- and subsequently
\begin{eqnarray}
    \sum_{r=1}^\mathcal{Z}e^{ikr}r^{-\alpha} &\approx& (-ik)^{\alpha-1}\Gamma(1-\alpha)\nonumber\\ &&+\sum_{n=0}^\infty \frac{\zeta(\alpha-n)}{n!}(ik)^n,
\end{eqnarray}
leading $\omega_k \propto k^{\alpha-1}$ at the critical point with $1<\alpha<2$ (weak long-range regime )~\cite{Ganesh2022,Cevolani2016,Hauke2013,Eisert2013}. In Fig.~\ref{fig:dispersion}(a), we plot $\omega_k$ as a function of $k$, demonstrating the crossover from the short-range to the long-range behavior of the dispersion relation (Eq.~(\ref{eq:dispersion})) via exact numerical evaluation of $\tilde J_k$, which takes place at $k=k_{\mathcal{Z}}$. The region $k<k_\mathcal{Z}$ with a unit slope indicates the short-range regime of the Brillouin zone, while the $\mathcal{Z}$-independent $k>k_\mathcal{Z}$ region with a slope $\alpha-1$ indicates the long-range regime. This is consistent with the results for {\it true} long-range system ($Z\to\infty$) for which the short-range regime disappears as $k_{\mathcal{Z}}\rightarrow 0$.

Going back to the $k\mathcal{Z}\ll 1$ limit,  for large $\mathcal{Z}$,
\begin{eqnarray}\label{eq:qp_velocity_largeZ}
v_{\alpha,\mathcal{Z}}\approx\mathcal{Z}^\eta,
\end{eqnarray}
with the exponent $\eta$, given by\footnote{The $\alpha$-dependence of $\eta$  is the same for $\alpha=2$ also, although a different methodology has to be adopted for proving this.} 
\begin{eqnarray}\label{eq:eta}
    \eta=\begin{cases}
        2-\alpha, &1< \alpha < 2\\
        0, &2\leq \alpha 
        \end{cases}
\end{eqnarray}
distinguishing the weak long-range regime $(1<\alpha<2)$ and the effective short-range $(\alpha>2)$ regimes in conjunction with the findings of~\cite{Cevolani2016,Hauke2013,Eisert2013}.   The variation of $\eta$ against $\alpha$, as depicted in Fig~\ref{fig:dispersion}(b), supports Eq.~(\ref{eq:eta}), where $\eta$ is calculated by exact evaluation of $d\omega_k/dk$ at the critical point for different $\log_{10}\mathcal{Z}$, followed by a numerical fit to a first degree polynomial. Note that $\eta=0$ beyond $\alpha=2$ indicates a $\mathcal{Z}$-independent $v_{\alpha,\mathcal{Z}}$ despite $\mathcal{Z}>1$, thereby indicating the $\alpha>2$ regime to be effective short-range. 

Notice that Eq.~(\ref{eq:qp_velocity_largeZ}) is consistent with the known limits  of the VREI model explored in the literature~\cite{PhysRevB.102.214203}. For $\mathcal{Z}=1$, one retrieves the well-known result for the quasi-particle velocity of the transverse field Ising model~\cite{Pfeuty1970}, which is independent of $\alpha$. For the $\mathcal{Z}\rightarrow\infty$, the maximum group velocity becomes a function of the power-law exponent $\alpha$, and can be obtained from the small-$k$ behavior of the dispersion. In the $\alpha<2$ regime, the maximum group velocity can, in principle, diverge when $k\to 0$ \cite{PhysRevB.102.214203,Vodola2015}, 
\begin{equation}
    v_\text{max}\sim \lim_{k\to 0}\frac{d\omega_k}{dk} \sim \lim_{k\to 0} k^{\alpha-2} \to \infty .
\end{equation}

\subsection{Correlation functions}
\label{subsec:correlations}

For finite $\mathcal{Z}>1$, the short-range (long-range) regime of the Brillouin zone is given by $k\ll k_\mathcal{Z}$ ($k\gg k_\mathcal{Z}$), corresponding to $r\gg \pi/k_\mathcal{Z}=\mathcal{Z}$ ($r\ll \mathcal{Z}$) in the real space. Since correlations (including quantum correlations belonging to the entanglement-separability as well as information-theoretic paradigms -- bipartite as well as  multipartite~\cite{Amico2008,Vidal2003,Titas2016,Pezze2017}) can be decomposed to two-point fermionic correlation functions $\alpha_r = \langle c_r c_0^\dagger \rangle$ and $\beta_r = \langle c_r c_0 \rangle$, we calculate $\alpha_r$ and $\beta_r$ in the thermodynamic limit from the Eq.~(\ref{eq:eigenstates}) as
\begin{eqnarray}\label{eq:alphar_betar}
    \alpha_r &=& \langle c_r c_0^\dagger \rangle = \frac{1}{\pi} \int_{0}^\pi dk |U_k|^2 \cos{kr} , \\
    \beta_r &=& \langle c_r c_0 \rangle = \frac{1}{\pi} \int_{0}^\pi dkU_k V_k^* \sin kr,    
\end{eqnarray}
with 
\begin{eqnarray}\label{eq:single_site}
    \alpha_0=\frac{1}{\pi}\int_0^\pi |U_k|^2dk
\end{eqnarray}
describing the special case of a single site $(r=0)$. For the $\mathcal{Z}=1$ limit, i.e., the transverse-field Ising model,  the short-range correlation length~\cite{Fisher1964-gw,Sachdev2011-jw}
\begin{equation}
    \xi_S=|\Delta h|^{-1},
\end{equation}
where $\Delta h=h-h_c$ quantifies the distance of the chosen value from the critical point $h_c$ on the $h$ axis. The dominant correlation $\beta_r$\footnote{In the $\mathcal{Z}\rightarrow\infty$ limit, $\alpha_r$ is shown to decay faster than $\beta_r$ when $|\Delta h|>0$ for the entire range $1<\alpha<2$~\cite{Debasis2021},  same as the $\mathcal{Z}=1$ limit, which is similar to the short-range regime $\alpha>2$. Our numerical investigation also confirms that $\beta_r$ is indeed the dominant correlation in the entire parameter regime considered in this paper.} is known to satisfy~\cite{Sachdev2011-jw,Pfeuty1970}
\begin{eqnarray}\label{eq:short_range_corr_decay}
    \beta_r\sim\begin{cases}
        r^{-1}, & r\ll\xi_S\\
        e^{-r/\xi_S}, & r\gg \xi_S
    \end{cases}
\end{eqnarray}
while in the true long-range limit $\mathcal{Z}\rightarrow \infty$ a long-range correlation length 
\begin{equation}
    \xi_L=|\Delta h|^{-1/(\alpha-1)},
\end{equation}
is shown to exist for $1<\alpha<2$ ~\cite{Debasis2021},  and a different behavior of the dominant correlation, given by
\begin{eqnarray}\label{eq:long_range_corr_decay}
    \beta_r\sim\begin{cases}
        r^{-1}, & r\ll\xi_L\\
        r^{-\alpha}, & r\gg \xi_L
    \end{cases}
\end{eqnarray}
is reported. Fig.~\ref{fig:dispersion}(c) shows the short-range regime $r>\mathcal{Z}$ in the thermodynamic limit, where the algebraic decay in the $r\ll \xi_S$ regime changes over to an exponential decay for $r\gg \xi_S$, which is consistent with the Eq.~(\ref{eq:short_range_corr_decay}). On the other hand, in Fig.~\ref{fig:dispersion}(d), we demonstrate, for the same model, the existence of long-range behavior in the correlations when $r<\mathcal{Z}$. The two algebraic decay regions $r\ll\xi_L$ and $r\gg\xi_L$ with different decay exponents given by Eq.~(\ref{eq:long_range_corr_decay}) can be seen from the figure.

\begin{figure}
    \centering
    \includegraphics[width=0.7\linewidth]{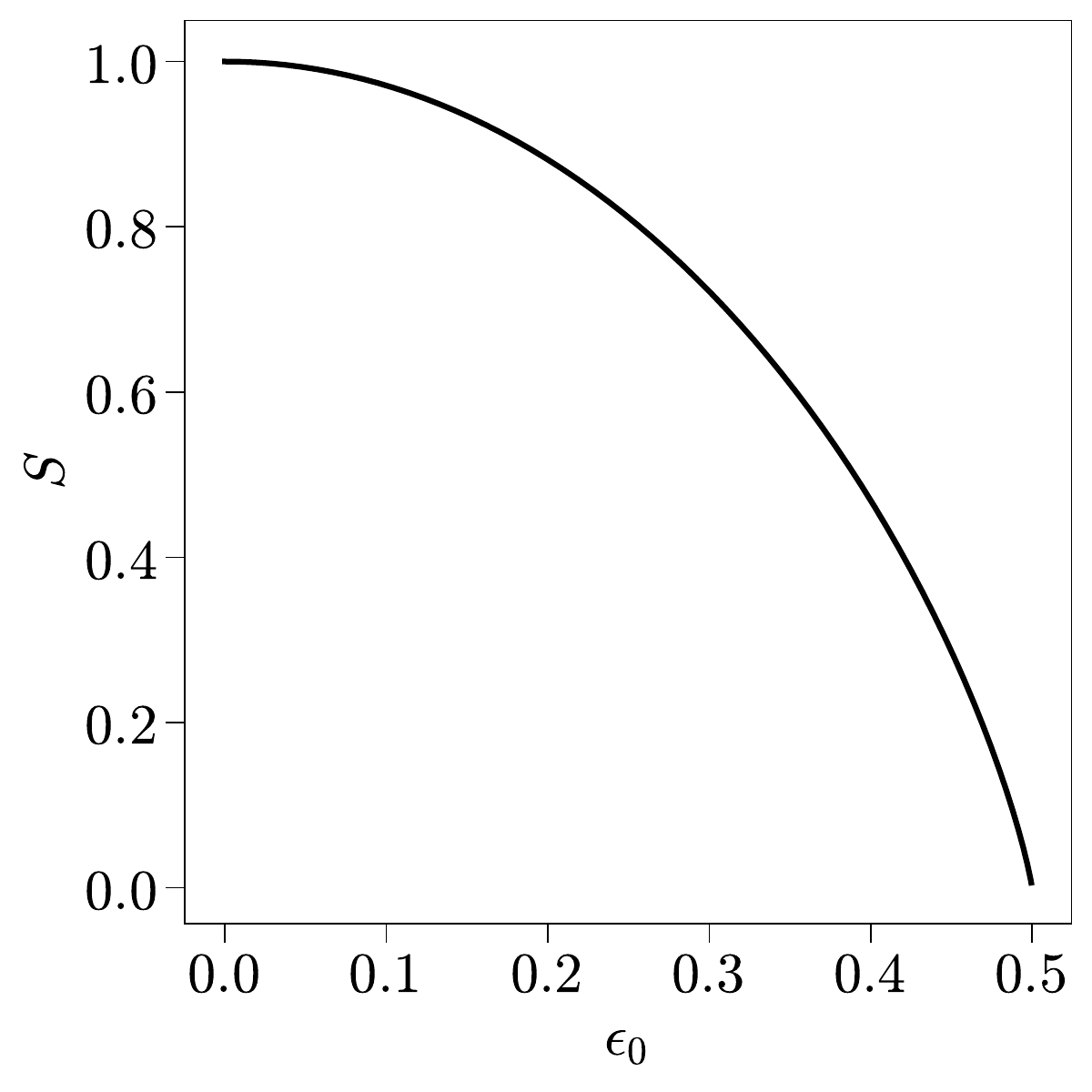}
    \caption{Variation of $S$ as a function of $\epsilon_0$ (see Eq.~(\ref{eq:entropy}))}
    \label{fig:S_vs_eps}
\end{figure}

\section{Scaling of Bipartite entanglement}
\label{sec:biartite_entanglement}

\begin{figure*}
    \centering
    \includegraphics[width=0.7\linewidth]{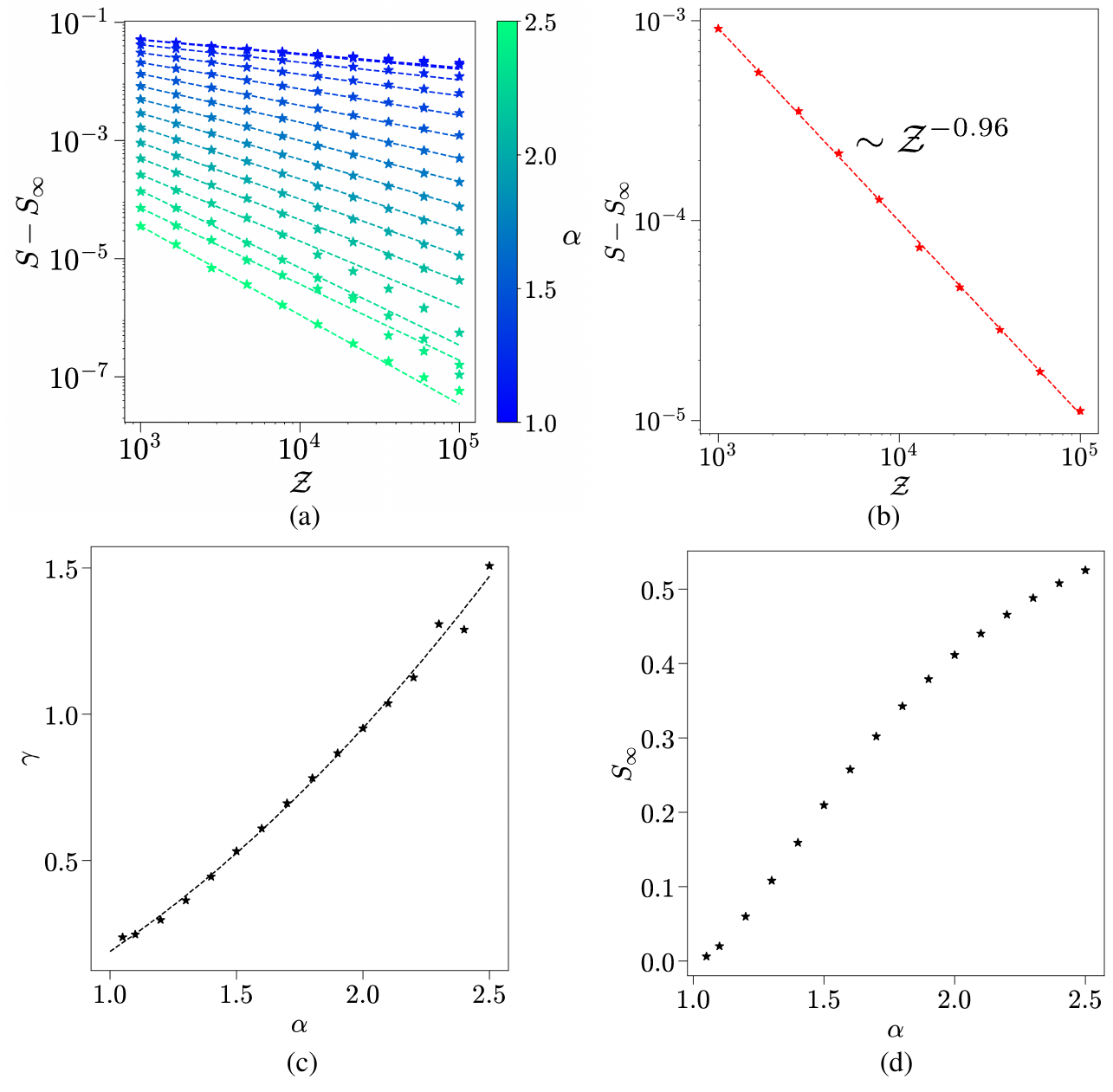}
    \caption{(a) Variation of $S-S_\infty$ as a function of $\mathcal{Z}$ with $M=1$ for different values of $\alpha$ in the range $1.0<\alpha<2.5$. The dashed straight lines correspond to fitting the data for $\mathcal{Z}<10^4$ to Eq.~(\ref{eq:entropy_M}), where $\gamma$ and $S_\infty$ are given in (c) and (d) respectively. (b) Plot of $S-S_\infty$ against $\mathcal{Z}$ for the specific case of $\alpha=2.0$, where $\gamma\approx 1$, which matches well with our theoretically predicted value in Eq.~(\ref{eq:alpha_2_result}). The deviation of $\gamma$ from $\gamma=1$ is clearly observed in (a) for values of $\alpha$ away from $2$. (c) Behavior of  $\gamma$ (see Eq.~(\ref{eq:entropy_M})), as obtained by the slopes of the fitted straight lines in (a), as a function of $\alpha$. The dotted line indicates the numerical fit to the Eq.~(\ref{eq:gamma_alpha}) with $c_0=-0.20(4),c_1=0.2(1)$ and $c_2=0.1(1)$. (d) Variation of  $S_\infty$ (see Eq.~(\ref{eq:entropy_M})) with $\alpha$ in the limit $\mathcal{Z} \rightarrow \infty$.}
    \label{fig:static_entropy}
\end{figure*}

We now explore how the bipartite entanglement between a block of $M$ qubits and the rest of the system varies with $\mathcal{Z}$. The bipartite entanglement between an $M$-qubit block and the rest of the system is quantified by the von Neumann entropy of the reduced state $\rho_M=\text{Tr}_{\overline{M}}\rho$ of the $M$-qubit block, given by 
\begin{equation}\label{eq:entropy_original}
    S=-\sum_i\lambda_i\log_2\lambda_i,
\end{equation}
where $\rho$ is the state of the full system, $\lambda_i$ ($i=1,2,\cdots,2^M$) are the eigenvalues of $\rho_M$, and the partial trace in calculating $\rho_M$ is taken over all qubits outside $M$. For the sake of discussion, let us first consider the  special case of $M=1$, for which the single-site density matrix $\rho_i$ corresponding to the site $i$ is given by  $\rho_i=\text{diag}\{\alpha_0,1-\alpha_0\}$ $\forall i$ (due to the translation symmetry), leading to
\begin{eqnarray}\label{eq:binary_entropy}
    S=-\alpha_0\log_2\alpha_0-(1-\alpha_0)\log_2(1-\alpha_0),
\end{eqnarray}
where $0\leq \alpha_0\leq 1$ (see Eq.~(\ref{eq:single_site})). By introducing a new variable $\epsilon_0=\alpha_0-1/2$, $S$ can be rewritten as an even function of $\epsilon_0$ (see Appendix~\ref{app:entropy_calculation} for details): 
\begin{equation}\label{eq:entropy}
    S=1-\frac{1}{2}\log_2(1-4\epsilon_0^2)-\epsilon_0\log_2\bigg(\frac{1+2\epsilon_0}{1-2\epsilon_0}\bigg). 
\end{equation}
A series expansion, as discussed in Appendix~\ref{app:entropy_calculation}, shows that $S$ is a monotonically decreasing function of $\epsilon_0^2$ (see Fig.~\ref{fig:S_vs_eps}).

\begin{figure*}
    \centering
    \includegraphics[width=0.9\linewidth]{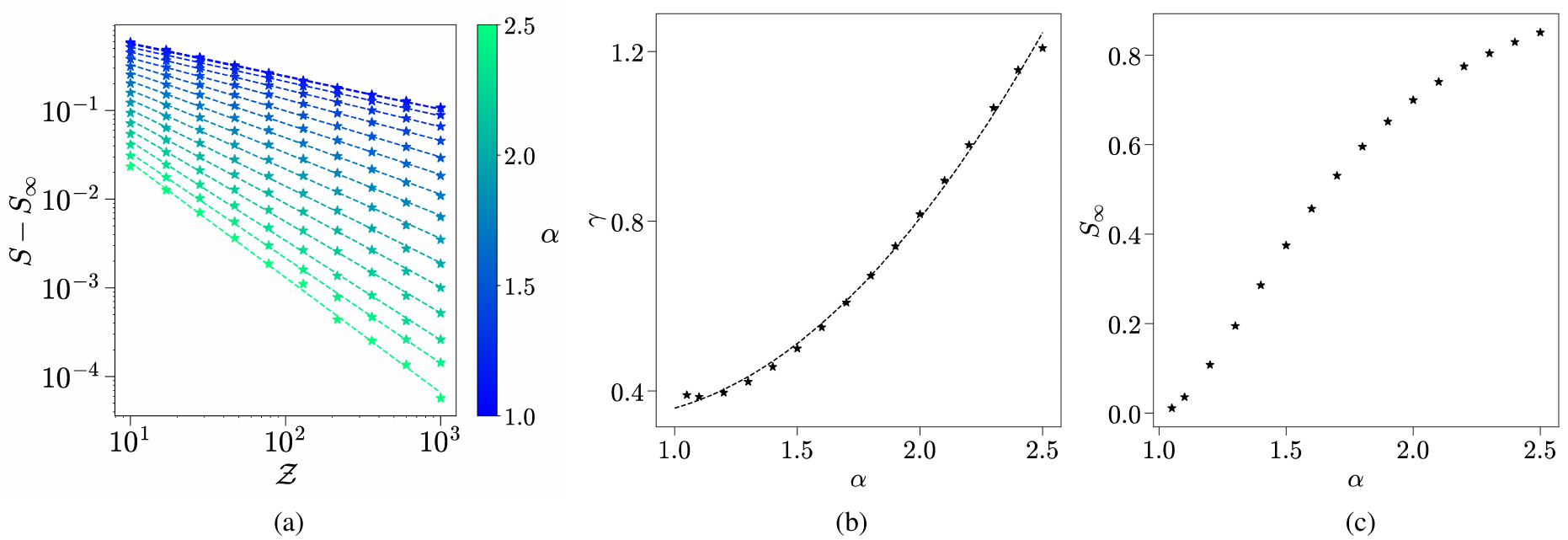}
    \caption{(a) Variation of $S-S_\infty$ as a function of $\mathcal{Z}$ for the case of $M=4$ with $1<\alpha<2.5$. Dashed lines are obtained by fitting the data corresponding to each $\alpha$. (b) Behavior of  $\gamma$ (see Eq.~(\ref{eq:entropy_M})), as obtained by the slopes of the fitted straight lines in (a), as a function of $\alpha$. The dotted line indicates the numerical fit to the Eq.~(\ref{eq:gamma_alpha}) with $c_0=0.48(2),c_1=-0.40(7)$ and $c_2=0.28(6)$. (c) Variation of  $S_\infty$ (see Eq.~(\ref{eq:entropy_M})) with $\alpha$ in the limit $\mathcal{Z} \rightarrow \infty$ for $M=4$.}
    \label{fig:static_entropy2}
\end{figure*}

We now compute $\alpha_0$ (i.e., $\epsilon_0$), and due to the natural breakup of Brillouin zone at $k=k_{\mathcal{Z}}=\pi/\mathcal{Z}$ into a short-range and a long-range regimes (see Sec.~\ref{sec:model_and_diagonalization}), we evaluate Eq.~(\ref{eq:single_site}) using the trapezoidal rule, where the integral is approximated using trapeziums constituted of the points $|U_0|^2$, $|U_{\pi/\mathcal{Z}}|^2$, and $|U_\pi|^2$ (see Appendix~\ref{app:entropy_calculation} for details).  This leads to
\begin{eqnarray}\label{eq:epsilon_0}
    \epsilon_0\approx \frac{C}{2}-\frac{1}{4\mathcal{Z}},
\end{eqnarray}
where
\begin{equation}\label{eq:C}
    C=\Bigg[1+\bigg(\frac{g_{\pi/\mathcal{Z}}}{1+\sqrt{1+g_{\pi/\mathcal{Z}}^2}}\bigg)^2\Bigg]^{-1}
\end{equation}
becomes independent of $\mathcal{Z}$ in the limit $\mathcal{Z}\rightarrow\infty$ as $g_{\pi/\mathcal{Z}}$ has the limiting value
\begin{equation}\label{eq:g_pi_Z}
    \lim_{\mathcal{Z}\rightarrow\infty}g_{\pi/\mathcal{Z}}=\frac{6-2\alpha}{\pi(2-\alpha)}. 
\end{equation}
Further, to the leading order in $\epsilon_0$, $S\approx1-2\log_2(\text{e})\epsilon_0^2$,  and subsequently (using Eq.~(\ref{eq:epsilon_0}))
\begin{equation}\label{eq:alpha_2_result}
    S\approx S_\infty+\kappa\mathcal{Z}^{-1}
\end{equation}
to the leading order in $\mathcal{Z}^{-1}$, 
where $S_\infty$ captures the saturation of block entropy in the limit $\mathcal{Z}\rightarrow \infty$, and $\kappa$ is a proportionality constant. This indeed indicates a decrease in entropy at the critical point with $\mathcal{Z}$, contrary to expectation. We perform extensive numerical analysis of the behavior of $S$ at the critical point with $M=1$ (see Fig.~\ref{fig:static_entropy}(a)), and find Eq.~(\ref{eq:alpha_2_result}) to be explaining our numerical data only in the case where $\alpha=2$ (see Fig.~\ref{fig:static_entropy}(b)). As $\alpha$ moves away from $2$, Eq.~(\ref{eq:alpha_2_result}) modifies to 
\begin{equation}\label{eq:entropy_M}
    S= S_\infty+\kappa\mathcal{Z}^{-\gamma(\alpha)},
\end{equation}
where the exponent $\gamma(\alpha)$ is a function of $\alpha$. However, this $\alpha$-dependence of the exponent is lost in the approximations made during the evaluation of $\alpha_0$. Our numerical analysis suggests a quadratic form for $\gamma(\alpha)$ as
\begin{equation}\label{eq:gamma_alpha}
    \gamma(\alpha)=c_0+c_1\alpha+c_2\alpha^2,
\end{equation}
where the constants $c_0$, $c_1$, and $c_2$ are estimated in Fig.~\ref{fig:static_entropy}(c). The variation of $S_\infty$ as a function of $\alpha$ is included in Fig.~\ref{fig:static_entropy}(d).

Note that lack of existence of closed forms for $|U_k|^2$  motivated us to use numerical integration techniques for evaluating $|U_k|^2$ exactly (using \href{http://www.gnu.org/software/gsl/}{GSL} in C++), which is required for computing $S$. We point out here  that $\tilde{J}_k$ (Eq.~(\textcolor{red}{12}), and subsequently $U_k$ (Eq.~(\textcolor{red}{15})) exhibit highly oscillatory behaviour in $k$ as $\mathcal{Z}$ increases, leading to large deviations from their actual values even when numerical techniques are used, compared to the well-behaved case of small $\mathcal{Z}$. Also, for optimally using computational resource, one needs to set a cut-off in the  absolute error occurring in the estimated correlation functions thereby setting the stopping criteria for the numerical integration algorithm. These factors, when combined with the fixed floating-point precision of the computer, result in larger errors in the estimated correlation function for larger values of $\mathcal{Z}$, which get propagated non-linearly to the value of $S$ due to Eq.~(\ref{eq:entropy_original}), and are manifested in the large $\mathcal{Z}$ regime of Fig.~\ref{fig:static_entropy}(a) by the data deviating from the expected analytical trend.

For $M>1$, $\lambda_i$'s can be calculated as the eigenvalues of the correlation matrix $\Pi(M)$ of size $2M\times2M$~\cite{Vidal2003,Eisert2010,Amico2008}, defined as 
\begin{equation}
    \Pi(M)=\begin{bmatrix}
        1-G & F\\
        F^\dagger & G
    \end{bmatrix},
    \label{eq:correlation_matrix}
\end{equation}
where  $G$ and $F$ are the fermionic correlation matrices corresponding to two sites $i$ and $j$ at a distance $r=|i-j|$, given by 
\begin{eqnarray}
    G_{i,j}&=&\alpha_{i-j},\\
    F_{i,j}&=&\beta_{i-j}.
\end{eqnarray}
While calculation of $G$ and $F$ for arbitrary $r$ becomes extensive with increasing $M$, approximations similar to the case of $M=1$ can still be carried out, and a decrease in $S$ with increasing $\mathcal{Z}$ at the critical point is expected. Also, for arbitrary $M$, $S$ has contributions from both the long-range $(r<\mathcal{Z})$ as well as the short-range $(r>\mathcal{Z})$ correlations,  and the quasi-particle velocity plays a crucial role in spreading the correlations across the system. Since the long-range regime is always finite, one expects the major contribution to the correlations, and subsequently $S$ to be due to the short-range quasi-particle velocity, the asymptotic form of which  is given in Eq.~(\ref{eq:qp_velocity_largeZ}) at the critical point. With these insights, we propose Eq.~(\ref{eq:entropy_M}) as the asymptotic form for block entropy in the large $\mathcal{Z}$ limit for arbitrary $M$, where the constants $S_\infty,\kappa$ and $\gamma$ are functions of $M$ and $\alpha$. 

We perform extensive numerical  analysis of the behavior of $S$ at the critical point with varying $\mathcal{Z}$ in the thermodynamic limit. In Fig.~\ref{fig:static_entropy2}(a) we present the log-log plot $S-S_\infty$ against $\mathcal{Z}$ for $M=4$, which establishes the validity of Eq.~(\ref{eq:entropy_M}) also for $M>1$, while our numerical analysis up to $M=100$ reveals it to remain the same irrespective of choice of $M$. The variations of $\gamma$ and $S_\infty$ against $\alpha$ are similar to those in the cases of $M=1$, and are shown in Fig.~\ref{fig:static_entropy2}(b) and Fig.~\ref{fig:static_entropy2}(c) respectively. Note that the discussion on the possible sources of errors in numerical calculation of $S$ preceding Eq.~(\ref{eq:correlation_matrix}) is valid for $M>1$ also. Due to small values of $S-S_\infty$, and subsequent increase in the numerical errors, we restrict the sampling to be in the regime $1< \alpha<2.5$ and $\mathcal{Z}\leq10^3$ (see Fig.~\ref{fig:static_entropy2}(a)). However, in order for getting a closer-to-actual value of $\gamma$, one may need to explore larger $\mathcal{Z}$ regime, where an increase in noise in the numerical data is expected.

Note that $\gamma\rightarrow1$ as $\alpha\rightarrow2$ for different values of $M$. The exponent $\gamma>1$ in the region $\alpha>2$ indicates a rapid decay of $S$ towards $S_\infty$, and making the bipartite entanglement practically independent of $\mathcal{Z}$. This is in clear agreement with the $\mathcal{Z}$-dependence of the quasi-particle velocity,  and distinguishes the weak long-range regime  $(1<\alpha<2)$ from the effective short-range regime$(\alpha>2)$~\cite{Cevolani2016,Hauke2013,Eisert2013}.

\begin{figure*}
    \centering
    \includegraphics[width=0.7\linewidth]{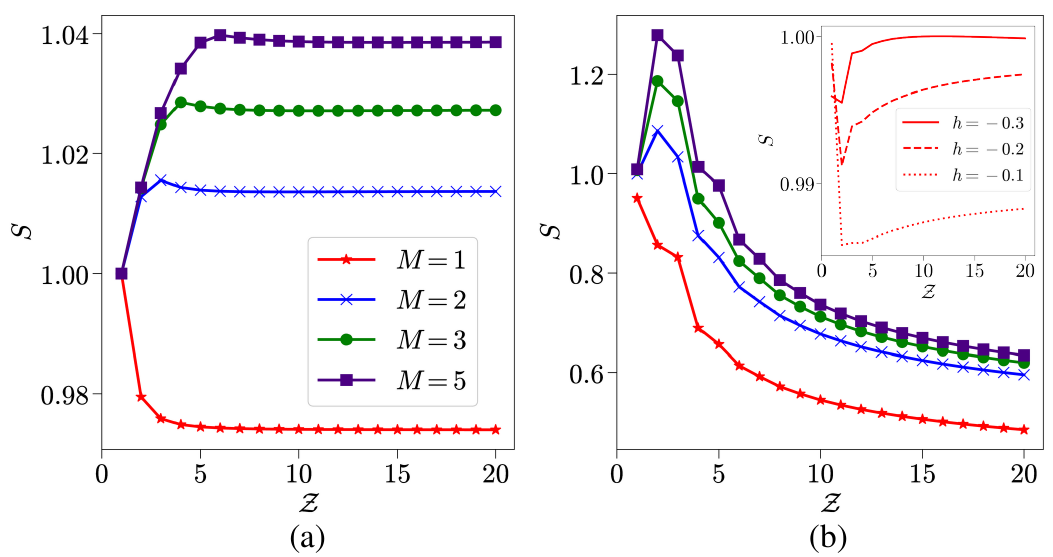}
    \caption{Variation of $S$ with $\mathcal{Z}$ at the off-critical points (a) $h=0$, and (b) $h=-1$ for different values of $M$. The inset of (b) shows the variation of $S$ with $\mathcal{Z}$ in the case of $M=1$ for three different values of $h$. For demonstration, we have chosen $\alpha=1.5$.}
    \label{fig:off_critical}
\end{figure*}

So far, the calculations presented are specifically at the critical point $h_c=2$. For the off-critical points ($h_c\neq 2$), analytical calculations become cumbersome even for the simplest case of $M=1$ due to the complex forms of $U_k$ and $V_k$  (Eq.~(\ref{eq:eigenstates})). Nevertheless,  a numerical investigation into the behaviour of $S$ is still possible. For example, the variations of $S$ with $\mathcal{Z}$ for different values of $M$ at the off-critical points $h=0$ and $h=-1$ are depicted in Fig.~\ref{fig:off_critical}(a) and (b). While $S$ exhibits an initial decrease followed by a saturation with $\mathcal{Z}$ for $M=1$, with $M>1$, $S$ increases with $\mathcal{Z}$ before attaining a saturation. Our numerical analysis reveals that for $M=1$, this trend of decreasing $S$ with $\mathcal{Z}$ persists for all values of $h$ except in a small interval of $h$ less than $h=0$ (see inset of Fig.~\ref{fig:off_critical}(b)) where there is a nominal increase before attaining the saturation can be observed. 

Note further that the observed power-law decay of the bipartite entanglement with $\mathcal{Z}$ at the fixed critical point $h_c=2$  is specific to the VREI model, and may not be present in other variable-range Hamiltonians. For example, consider the variable-range Ising (VRI) Hamiltonian~\cite{Koffel2012,Vodola2015}, where the interaction terms differ from that of VREI via an absence of $\prod_m\sigma_m^z$, while other system-parameter of the VRI Hamiltonian being the same as those of the VREI Hamiltonian, making the latter non-integrable. A numerical analysis of the finite-sized VRI model reveals an increase in $S$ as one increases $\mathcal{Z}$, which is in contrast to the result reported in this paper, thereby making our result specific for the in the VREI model.

\section{Dynamics of bipartite entanglement}
\label{sec:dynamics}

We now consider a situation where the system is initialized ($t=0$) in the \emph{fully separable} ground state $\ket{G}=\ket{0}^{\otimes N}$ of $H$ in the limit $h\rightarrow\infty$. At $t>0$, the field is suddenly quenched to the critical value $h=2$, such that the system undergoes a time-evolution given by $\ket{\psi(t)}=\exp -i Ht\ket{G}$. In the following, we explore transient as well as steady state bipartite entanglement between a block of qubits of size $M$  and the rest of the qubits vary with $\mathcal{Z}$.

\subsection{Transient entanglement growth}

First we look at the transient growth regime of $S(t)$ for small $t$. It has been shown, for $\mathcal{Z}=1$, that $S(t)$ grows linearly with $t$ up to the time $\tau=M/2v$, and saturates thereafter to a value $S(t\geq \tau)=S_{\text{sat}}$ proportional to $M$~\cite{Calabrese2005}, where $M$ is the block size and $v$ is the quasi-particle velocity. In the case of VREI model, $S_{\text{sat}}$ decreases with $\mathcal{Z}$ for a given value of $M$, and given that 
$v_{\alpha,\mathcal{Z}}=\mathcal{Z}^\eta$ for $\mathcal{Z}\gg 1$ (see Eq.~(\ref{eq:qp_velocity_largeZ})), the time $t$ taken to achieve the saturation is expected to satisfy $t\gg \tau_{\alpha,\mathcal{Z}}$, where 
\begin{equation}\label{eq:tau}
    \tau_{\alpha,\mathcal{Z}}=\frac{M}{2v_{\alpha,\mathcal{Z}}}\propto\mathcal{Z}^{-\eta}.
\end{equation}
For $t\ll\tau_{\alpha,\mathcal{Z}}$, due to the integrability and existence of a quasi-particle picture as in~\cite{Calabrese2005}, $S(t)$ is expected to be linear in $t$. Note that the quasi-particle velocity (Eq.~(\ref{eq:qp_velocity_largeZ})) is valid for $r\gg\mathcal{Z}$, implying that the effect of $\tau_{\alpha,\mathcal{Z}}$ (Eq.~(\ref{eq:tau})) is prominent for  $M\gg\mathcal{Z}$, and is less evident for $M\ll\mathcal{Z}$ (for example, $M=1$, as shown in Fig.~\ref{fig:dynamics}(b)) due to the diverging velocities in the effective long range regime. We demonstrate $\tau_{\alpha,\mathcal{Z}}$ in Fig.~\ref{fig:dynamics}(a), where the variation of $S$ against $t/\tau_{\alpha,\mathcal{Z}}$ is shown for different values of $\mathcal{Z}$ for $M=50$. Note, in Fig.~\ref{fig:dynamics}(a), that $S$ linearly increases with $t$ for $t<\tau_{\alpha,\mathcal{Z}}$, and saturates to a finite value $S_{\text{sat}}$ that decreases with $\mathcal{Z}$.

\subsection{Steady state entanglement}
\begin{figure*}[ht]
    \centering
    \includegraphics[width=0.8\linewidth]{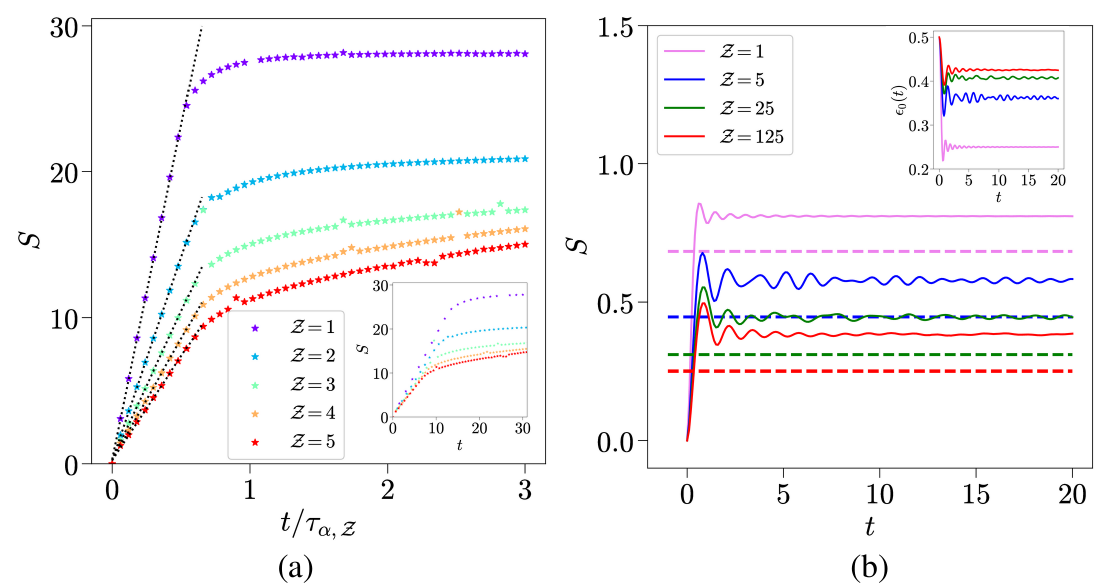}
    \caption{(a) Variation of $S$ against $t/\tau_{\alpha,\mathcal{Z}}$ for different values of $\mathcal{Z}$ with $M=50$ and $\alpha=1.5$. The black dotted straight lines are fitted using the first eight data points for $t<\tau_{\alpha,\mathcal{Z}}$, demonstrating the linear growth of $S(t)$. The inset depicts the same data with $t$ without the scale $\tau_{\alpha,\mathcal{Z}}$. (b) Variation of $S$ against $t$ for different values of $\mathcal{Z}$ with $M=1$ and $\alpha=1.5$. The entropy attains a steady state value $>S^{\text{static}}$, shown by the horizontal dashed lines for different $\mathcal{Z}$. The inset exhibits the variation of $\epsilon_0(t)$ for the same parameter values.}
    \label{fig:dynamics}
\end{figure*}

We now investigate the system after attaining the steady state as $t\rightarrow\infty$, where, $S(t)$ exhibits small fluctuations about a mean value (see Fig.~\ref{fig:dynamics}(b)). We consider the long-time average of $S$, given by 
\begin{equation}
    \overline{S}=\frac{1}{T}\int_{t_0}^{t_0+T}Sdt.
\end{equation}
Here, $t_0$ is chosen to be a time at which the system has already attained the steady-state, and $T$ is the time period over which the long-time average is computed.

The initial product state corresponds to the zero particle state in the momentum modes $(c_k^\dagger c_k=0)$, which evolves in time as
\begin{equation}
    \ket{\psi_k(t)}=\begin{bmatrix}
        u_k(t)\\ v_k(t)
    \end{bmatrix}=\begin{bmatrix}
        U_k^2\text{e}^{i\omega_kt}+V_k^2\text{e}^{-i\omega_kt}\\
        U_k V_k(\text{e}^{i\omega_kt}-\text{e}^{-i\omega_kt})
    \end{bmatrix},
\end{equation}
were $U_k,V_k,\omega_k$ are given by Eqs.~(\ref{eq:eigenstates}) and (\ref{eq:dispersion}) respectively. Algebraic simplification leads to 
\begin{equation}
    |u_k(t)|^2=1-4\sin^2(\omega_kt)U_k^2V_k^2,
\end{equation}
followed by the use of Eq.~(\ref{eq:single_site}), resulting in 
\begin{eqnarray}
    \epsilon_0(t)&=&\frac{1}{\pi}\int_0^\pi |u_k(t)|^2dk-\frac{1}{2},\nonumber\\
    &=&\frac{1}{2}-\frac{4}{\pi}\int_0^\pi \sin^2(\omega_kt)|U_k|^2dk\nonumber\\&&+\frac{4}{\pi}\int_0^\pi \sin^2(\omega_kt)|U_k|^4dk,
\end{eqnarray}
where we have used $V_k^2=1-U_k^2$. We proceed by evaluating the integrals approximately using the trapezoidal rule, as in the static case (see Sec.~\ref{sec:biartite_entanglement} and Appendix~\ref{app:entropy_calculation}). Noticing that the time-dependent sinusoidal part integrates to $1/2$ in the long-time limit, the time averaged value  of $\epsilon_0$, defined by
\begin{eqnarray}
    \overline{\epsilon_0}=T^{-1}\int_{t_0}^{t_0+T}\epsilon_0(t)dt,
\end{eqnarray}
becomes 
\begin{equation}
    \overline{\epsilon_0}=\frac{1}{2}-\frac{2}{\pi}\int_0^\pi |U_k|^2dk+\frac{2}{\pi}\int_0^\pi |U_k|^4dk,
\end{equation}
where the second term on the R.H.S has  already been evaluated (see Sec.~\ref{sec:biartite_entanglement}). The third term on the R.H.S. can also be evaluated in the same fashion using $|U_{\pi/\mathcal{Z}}|^4=C^2$, where $C$ is given by Eq.~(\ref{eq:C}). This leads to 
\begin{equation}\label{eq:quenched_entropy}
    |\overline{\epsilon_0}|=\epsilon_0^{\text{static}}-\bigg(\frac{3}{2}C-C^2-\frac{1}{2}-\frac{1}{2\mathcal{Z}}\bigg),
\end{equation}
where $\epsilon_0^{\text{static}}$ is the value of $\epsilon_0$ obtained in the static case (see Eq.~(\ref{eq:epsilon_0})). Since $1/2\leq C \leq 1$, the quantity within the parenthesis in the R.H.S. of Eq.~(\ref{eq:quenched_entropy}) is always positive ensuring $\overline{S}>S^{\text{static}}$, where $S^{\text{static}}$ is the value of $S$ in the ground state of the VREI model at the critical point. In Fig.~\ref{fig:dynamics}(b), we demonstrate this explicitly for the case of $M=1$. Further, by virtue of Eqs.~(\ref{eq:quenched_entropy}) and (\ref{eq:epsilon_0}), $\overline{S}$ exhibits a similar behavior against $\mathcal{Z}$ as in the static case,  with $\overline{S}\propto \mathcal{Z}^{-\gamma}$, although the details of the dependence of $\gamma$ on $\alpha$ would differ.

\section{Conclusion}
\label{sec:conclusion}

In summary, we investigate the two-point correlation functions and bipartite entanglement between a block of $M$ qubits with the rest in the ground state of the one-dimensional variable-range extended Ising model. Here, the variable range of interaction is achieved by varying the coordination number, $\mathcal{Z}$, of each qubit, where the interaction strength between any two qubits at a distance $r$ varies as $\sim r^{-\alpha}$ such that the sum of all such interactions for a fixed $\mathcal{Z}$ adds up to $1$. Our results indicate that in the weak long-range regime,  $r<{\cal Z}$ exhibits long-range behavior of the correlation functions for any finite $\mathcal{Z}$, while the $r>{\cal Z}$ region behaves as an effective short-range regime. Further, we calculate the bipartite entanglement between a block of $M$ qubits with the rest of the system in the ground state of the Hamiltonian, as quantified by the von Neumann entropy, $S$ and estimate analytically that for $M=1$, $S\sim\mathcal{Z}^{-1}$. We perform extensive numerical analysis and find that $S\sim\mathcal{Z}^{-1}$ only for $\alpha=2$ in the case of $M=1$, while for other values of $\alpha>1$, $S\sim\mathcal{Z}^{-\gamma}$, with $\gamma$ being a quadratic function of $\alpha$. We numerically demonstrate the qualitative validity of $S\sim\mathcal{Z}^{-\gamma}$ and the related results in the case of arbitrary values of $M>1$ in the range $\alpha>1$. We further consider a dynamics of the system, initiated at the ground state of the infinite-field limit of the Hamiltonian and subsequently subjected to a sudden quench via the critical Hamiltonian. We show that the bipartite entanglement grows with time and eventually saturates with small oscillations about a steady mean value. The long-time average of bipartite entanglement, $\overline{S}$, exhibits a similar dependence on $\mathcal{Z}$ as in the static case, i.e., $\overline{S}\sim\mathcal{S}^{-\gamma}$, which we prove analytically for $M=1$, and verify numerically for arbitrary $M$.  

Our work opens up a number of interesting and pertinent questions. It would be interesting to analyze the strong long-range regime to see if similar scalings of correlation functions as well as entanglement with $\mathcal{Z}$ are obtained. Also, given the existence of evidence for feeble distant two-site entanglement in the weak long-range region~\cite{Ganesh2022}, it would be interesting to explore the effect of increasing $\mathcal{Z}$ on genuine multipartite entanglement present in the ground state of the model in the strong long-range regime. Further, going beyond the variable-range extended Ising model, the  behavior of bipartite as well as multipartite entanglement with increasing coordination number in other exactly-solvable and yet differently classified models remains to be investigated.

We acknowledge the use of the C++ libraries (a) \href{http://www.gnu.org/software/gsl/}{GSL}  for a wide range of mathematical routines,  and (b) \href{https://github.com/titaschanda/QIClib}{QIClib} for general purpose quantum information processing and quantum computing.

\acknowledgements 

A.K.P acknowledges the support from the Anusandhan National Research Foundation (ANRF) of the Department of Science and Technology (DST), India, through the Core Research Grant (CRG) (File No. CRG/2023/001217, Sanction Date 16 May 2024). D.S. acknowledges support from the IoE seed grant
from Banaras Hindu University (Sanction no. IoE/Seed Grant (5th call)/2024-25/81866 dated 13 December 2024) and the YSRA scheme of DAE, BNRS (Government of India). H.K.J. acknowledges
the Prime Minister Research Fellowship Program, Government of India, for the financial support.

\appendix

\section{Quasi-particle velocity}
\label{app:calculation}

In order to evaluate $\tilde{J}_k$ (Eq.~(\ref{eq:Jk_tilde})), and subsequently the effective dispersion, first we approximate $A$ as
\begin{eqnarray}
    A&=&
    \sum_{r=1}^\infty r^{-\alpha}-\sum_{r=\mathcal{Z}+1}^\infty r^{-\alpha},\nonumber\\
    &\approx& \zeta(\alpha)-\int_{r=\mathcal{Z}}^\infty r^{-\alpha}dr\nonumber\\
    &=& \zeta(\alpha) + \frac{\mathcal{Z}^{1-\alpha}}{1-\alpha}, 
\end{eqnarray}
followed by 
\begin{eqnarray}
    \sum_{r=1}^\mathcal{Z}e^{ikr}r^{-\alpha} &=& \mathbf{Li}_\alpha(e^{ik})-\sum_{r=\mathcal{Z}+1}^\infty e^{ikr}r^{-\alpha},\nonumber\\
    &\approx& \mathbf{Li}_\alpha(e^{ik})-\int_{r=\mathcal{Z}}^\infty e^{ikr}r^{-\alpha}dr\label{eq:app-polylog}, 
\end{eqnarray}
where $\zeta(\alpha)$ and $\mathbf{Li}_\alpha(e^{ik})$ are the Riemann zeta function, and the polylogarithm functions, respectively. A substitution of $t=-ikr$ and $dr=idt/k$ in the Eq.~(\ref{eq:app-polylog}) leads to 
\begin{eqnarray}
    \int_{r=\mathcal{Z}}^\infty e^{ikr}r^{-\alpha}dr &=& (-ik)^{\alpha-1}\int_{-ik\mathcal{Z}}^\infty e^{-t}t^{-\alpha}dt,\nonumber\\
    &=& (-ik)^{\alpha-1} \Gamma(1-\alpha,-ik\mathcal{Z}),
\end{eqnarray}
where   $\Gamma(s,x)=\int_x^\infty e^{-t} t^{s-1}dt$ is the upper incomplete Gamma function. Further,  considering the asymptotic form of the polylogarithmic function~\cite{Debasis2021}, given by 
\begin{eqnarray}
    \mathbf{Li}_\alpha(e^{ik}) &=& (-ik)^{\alpha-1}\Gamma(1-\alpha)\nonumber\\&&+\sum_{n=0}^\infty \frac{\zeta(\alpha-n)}{n!}(ik)^n,
\end{eqnarray}
one obtains
\begin{eqnarray}\label{eq:app-shortrange}
    \sum_{r=1}^\mathcal{Z}e^{ikr}r^{-\alpha} &=&(-ik)^{\alpha-1}\gamma(1-\alpha,-ik\mathcal{Z})\nonumber\\&&+ \sum_{n=0}^\infty \frac{\zeta(\alpha-n)}{n!}(ik)^n, 
\end{eqnarray}
where $\gamma(1-\alpha,-ik\mathcal{Z})=\Gamma(1-\alpha)-\Gamma(1-\alpha,-ik\mathcal{Z})$ is the lower incomplete Gamma function, defined as $\gamma(s,x)=\int_0^x e^{-t}t^{s-1}dt$. In the short-range limit $k\mathcal{Z}\ll1$, up to leading order in $k$, the lower incomplete Gamma function can be approximated as 
\begin{eqnarray}
    \gamma(1-\alpha,-ik\mathcal{Z}) &=& \int_0^{-ik\mathcal{Z}} e^{-t} t^{-\alpha}dt,\nonumber\\
    &=& \int_0^{-ik\mathcal{Z}} \left(1-t+\frac{t^2}{2!}\cdots\right) t^{-\alpha}dt \nonumber\\
    &\approx& \int_0^{-ik\mathcal{Z}} t^{-\alpha}dt -\int_0^{-ik\mathcal{Z}} t^{1-\alpha}dt\nonumber \\
    &=& (-ik)^{1-\alpha}\left[\frac{\mathcal{Z}^{1-\alpha}}{1-\alpha}+ik\frac{\mathcal{Z}^{2-\alpha}}{2-\alpha}\right], 
\end{eqnarray}
while 
\begin{equation}
    \sum_{n=0}^\infty \frac{\zeta(\alpha-n)}{n!}(ik)^n\approx\zeta(\alpha)+ik\zeta(\alpha-1). 
\end{equation}
Substituting in Eq.~(\ref{eq:app-shortrange}),  one obtains
\begin{eqnarray}
    \sum_{r=1}^\mathcal{Z}e^{ikr}r^{-\alpha} &\approx& \zeta(\alpha)+\frac{\mathcal{Z}^{1-\alpha}}{1-\alpha}+ik\bigg[\zeta(\alpha-1)+\frac{\mathcal{Z}^{2-\alpha}}{2-\alpha}\bigg],\nonumber\\
\end{eqnarray}
which leads to 
\begin{eqnarray}
    \lim_{k\rightarrow0} \text{Re}(\tilde{J}_k)
    &=&\frac{1}{A}\left[ \frac{\mathcal{Z}^{1-\alpha}}{1-\alpha}+\zeta(\alpha)\right] = 1 \\
    \lim_{k\rightarrow0} \text{Im}(\tilde{J}_k)&=&\frac{1}{A}\text{Im}\left(\sum_{r=1}^\mathcal{Z}e^{ikr}r^{-\alpha}\right) = kv_{\alpha,\mathcal{Z}},
\end{eqnarray}
in the small $k$ limit, where
\begin{equation}
v_{\alpha,\mathcal{Z}}=\frac{\zeta(\alpha-1)+(2-\alpha)^{-1}\mathcal{Z}^{2-\alpha}}{\zeta(\alpha)+(1-\alpha)^{-1}\mathcal{Z}^{1-\alpha}}    
\end{equation}
is  the quasi-particle velocity at the critical point.

\section{Entropy for single-qubit density matrix}
\label{app:entropy_calculation}

For the single-qubit density matrix ($M=1$), von Neumann entropy is given by Eq.~(\ref{eq:binary_entropy}), where $\alpha_0$ $(0\leq \alpha_0\leq 1)$ is the probability to obtain $\ket{0}$ upon a measurement of $\sigma_i^z$ on the qubit. Defining $\alpha_0=\frac{1}{2}+\epsilon_0$, $S$ takes the form (while we use natural logarithm for all our calculations, results qualitatively remain independent of choice of the base of logarithm)
\begin{equation}\label{eq:S_in_epsilon_0}
    S=\frac{1}{2}\left[\ln4-\ln(1-4\epsilon_0^2)\right]-\epsilon_0\ln\bigg(\frac{1+2\epsilon_0}{1-2\epsilon_0}\bigg), 
\end{equation}
with $-1/2\leq \epsilon_0\leq 1/2$. Series-expanding the natural logarithms in Eq.~(\ref{eq:S_in_epsilon_0}), one obtains
\begin{equation}\label{eq:entropy_epsilon_form}
    S=\frac{\ln4}{2}-\bigg[2\epsilon_0^2+\frac{4}{3}\epsilon_0^4+\cdots \bigg],
\end{equation}
where the terms in the square bracket in the R.H.S. are all positive,  implying that $S$ decreases with increasing $\epsilon_0$. We point out here that choosing base $2$ for the logarithm makes the first term on the R.H.S. unity, which corresponds to the maximal entanglement when $\epsilon_0=0$, while any other base resulting in a positive yet different coefficient for each term in the square brackets, making no change in the qualitative result otherwise. 

\begin{figure}
    \centering
    \includegraphics[width=0.7\linewidth]{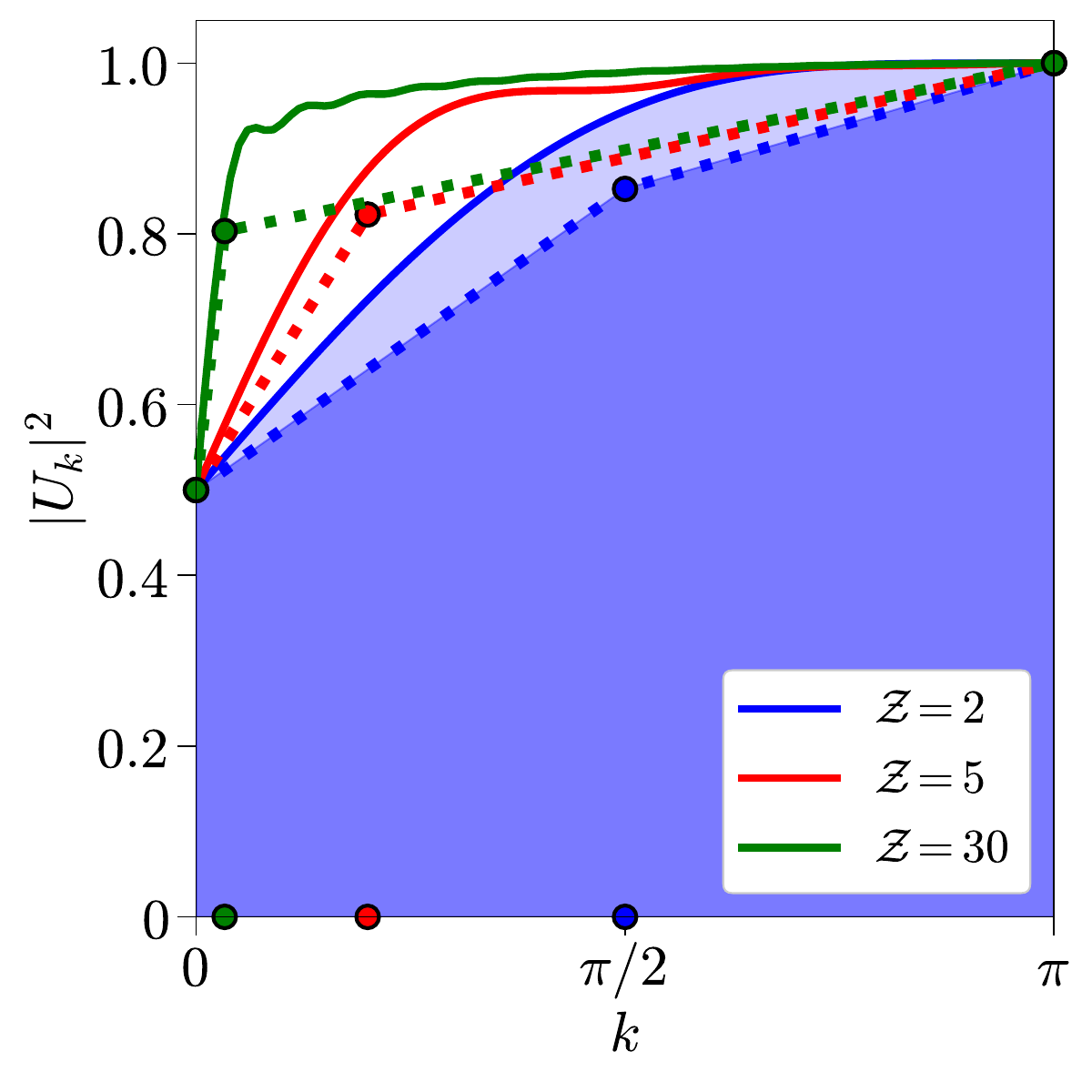}
    \caption{Variation of $|U_k|^2$ with $k$ for different $\mathcal{Z}$. The solid lines indicate exact values of $|U_k^2|$ obtained by normalizing Eq.~(\ref{eq:eigenstates}). We approximate area under $|U_k|^2$ (see, eg., the light shaded area in the case of $\mathcal{Z}=2$)  by the area under the colored dotted lines (the darker shaded area in the case of $\mathcal{Z}=2$). The points on the horizontal axis correspond to $k=\pi/\mathcal{Z}$ for the different $\mathcal{Z}$ considered, and the corresponding values of $|U_{\pi/\mathcal{Z}}|^2$ are shown on the dotted lines. All quantities plotted are dimensionless.}
    \label{fig:area_of_curve}
\end{figure}
Next, we evaluate $\alpha_0$, and hence $\epsilon_0$, at the critical point. For this, we divide the integral (Eq.~(\ref{eq:single_site})) into two parts according to the natural division of the Brillouin zone into the short-range ($k\in[0,\pi/\mathcal{Z}]$) and the long-range ($k\in[\pi/\mathcal{Z},\pi]$) domains, and approximate the area under each segment of the curve in the separate domains  using the trapezoidal rule  by using the points $|U_0|^2,|U_{\pi/\mathcal{Z}}|^2$ and $|U_\pi|^2$ on the curve (see Fig.~\ref{fig:area_of_curve} for a demonstration). To obtain these points, we note that
\begin{eqnarray}
    \tilde J_0&=&\frac{1}{A}\sum_r r^{-\alpha}=1,\nonumber\\
    \tilde J_\pi&=&\frac{1}{A}\sum_r (-1)^rr^{-\alpha}<1,
\end{eqnarray}
with $\text{Im}(\tilde J_0)=\text{Im}(\tilde J_\pi)=0$, which, when substituted in Eq.~(\ref{eq:eigenstates}), results in unnormalized $U_0,V_0,V_\pi=0$, and unnormalized $U_\pi>0$. Therefore, a normalization immediately yields
\begin{eqnarray}
    |U_0|^2&=&\frac{1}{2}, |U_\pi|^2=1.    
\end{eqnarray}
In order to obtain $|U_{\pi/\mathcal{Z}}|^2$ from the effective short-range limit $k\mathcal{Z}\ll1$, we evaluate $\gamma(1-\alpha,-ik\mathcal{Z})$, $\text{Re}(\tilde{J}_k)$, and $\text{Im}(\tilde{J}_k)$ up to the second order in $k$ as (see Eqs.~(\ref{eq:incomplete_gamma_simplified_first_order}),  and (\ref{eq:shortrangeReIm}) for expressions up to the first order in $k$)
\begin{eqnarray}\label{eq:incomplete_gamma_simplified}
    \gamma(1-\alpha,-ik\mathcal{Z}) &=&
    (-ik)^{1-\alpha}\bigg[\frac{\mathcal{Z}^{1-\alpha}}{1-\alpha}+ik\frac{\mathcal{Z}^{2-\alpha}}{2-\alpha}\nonumber\\&&-k^2\frac{\mathcal{Z}^{3-\alpha}}{6-2\alpha}\bigg],
\end{eqnarray}
and
\begin{eqnarray}\label{eq:Re}
    \text{Re}(\tilde{J}_k)&\approx&1-k^2R_{\alpha,\mathcal{Z}},\;
    \text{Im}(\tilde{J}_k)\approx k v_{\alpha,\mathcal{Z}},
\end{eqnarray}
with 
\begin{equation}\label{eq:F}
    R_{\alpha,\mathcal{Z}}=\frac{2^{-1}\zeta(\alpha-2)+(6-2\alpha)^{-1}\mathcal{Z}^{3-\alpha}}{\zeta(\alpha)+(1-\alpha)^{-1}\mathcal{Z}^{1-\alpha}},
\end{equation}
while $v_{\alpha,\mathcal{Z}}$ remains the same as in Eq.~(\ref{eq:qp_velocity}). Therefore, the normalized value of $|U_k|^2$ (using Eq.~(\ref{eq:eigenstates})) is 
\begin{equation}
    |U_k|^2=\Bigg[1+\bigg(\frac{g}{1+\sqrt{1+g^2}}\bigg)^2\Bigg]^{-1}
\end{equation}
where we have defined
\begin{equation}
    g=\frac{v_{\alpha,\mathcal{Z}}}{kR_{\alpha,\mathcal{Z}}}.
\end{equation}
In the large $\mathcal{Z}$ limit, $g(k=\pi/\mathcal{Z})$ is given by  Eq.~(\ref{eq:g_pi_Z}), implying that $|U_{\pi/\mathcal{Z}}|^2$ 
is a constant, $C$, that depends only on $\alpha$ (see Eq.~(\ref{eq:C})). Using these, one may now evaluate $\alpha_0$ as 
\begin{eqnarray}
    \alpha_0&=&\frac{1}{\pi}\int_0^{\pi/\mathcal{Z}} |U_k|^2dk + \frac{1}{\pi}\int_{\pi/\mathcal{Z}}^{\pi} |U_k|^2dk,\nonumber\\
    &\approx&\frac{1}{\pi}\bigg[\frac{\pi}{\mathcal{Z}}\bigg(\frac{|U_{\pi/\mathcal{Z}}|^2+|U_0|^2}{2}\bigg)\nonumber\\&&+\bigg(\pi-\frac{\pi}{\mathcal{Z}}\bigg)\bigg(\frac{|U_{\pi/\mathcal{Z}}|^2+|U_\pi|^2}{2}\bigg)\bigg],\nonumber\\
    &=&\frac{1}{2}+\bigg(\frac{C}{2}-\frac{1}{4\mathcal{Z}}\bigg),
\end{eqnarray}
where the term in the parenthesis can be identified to be $\epsilon_0$ (see Eq.~(\ref{eq:epsilon_0})). Since $C$ is a finite constant, $\epsilon_0$ increases monotonically with $\mathcal{Z}$ and saturates to a constant value indicating a decrease in entropy with $\mathcal{Z}$.

\bibliography{references}
 
\end{document}